\documentstyle{article}
\input tcilatex

\begin{document}

\title{QUASILUMPS FROM FIRST ORDER PHASE TRANSITIONS}
\author{}
\maketitle

\begin{abstract}
We investigated the bubble collisions during the first order phase
transitions. Numerical results indicate that within the certain range of
parameters the collision of two bubbles leads to formation of separate
relatively long-lived quasilumps - configurations filled with scalar field
oscillating around the true vacuum state. Energy is perfectly localized, and
density is slightly pulsating around its maximum. This process is
accompanied by radiation of scalar waves.
\end{abstract}

I. Dymnikova, L. Koziel \ {Institute of Mathematics, Informatics and Physics,%
\\University of Olsztyn, Zolnierska 14, 10-561 Olsztyn, Poland}

and M. Khlopov, S. Rubin \ Centre for Cosmoparticle Physics "Cosmion"\\%
Moscow State Engineering Physics Institute, Moscow, Russia



\vspace{0.2cm} ]

\section{Introduction}

Analysis of physical processes predicted in the early Universe on the basis
of particle theory, is the important way to study physical conditions in the
early Universe and physical mechanisms underlying those conditions. As a
result of such an analysis, the existence of hypothetical relics of early
Universe, such as primordial black holes, topologically stable or metastable
solitons etc, have been predicted. Confrontation of predicted effects with
observational data provides certain conclusions concerning both cosmological
evolution and particle physics models \cite{max}.

First order phase transitions as predicted by unified theories can occur at
several periods of cosmological evolution. They are predicted on the basis
of a wide class of models of particle symmetry breaking \cite{kolb}. They
are also considered as the final stage of inflation in a wide class of
inflationary models. Detailed study of nonlinear configurations arising at
the first order phase transitions and their dynamics is helpful not only for
cosmology - nonlinear dynamics of field theories describes a wide range of
phenomena occuring in laboratory physics.

The transition with symmetry breaking consists in decay of a metastable
phase by nucleation of bubbles of new phase \cite{Coleman}. A nucleated
bubble is a true vacuum fluctuation large enough to evolve classically. The
most likely fluctuation is a spherical bubble nucleated at rest with a
certain critical size determined by microphysical processes \cite{Coleman}.
Coleman \cite{Coleman} calculated the bubble nucleation rate in flat space
and at zero temperature using the euclidean path-integral formulation of a
scalar field theory. The nucleation rate in this case is proportional to $%
e^{-S_E}$ where $S_E$ is the euclidean action and solution to the euclidean
equation of motion for minimal action is the $O(4)$ symmetric "bounce"
solution.

In the very early Universe phase transitions most likely occur at a finite
temperature due to the fact that the form of a scalar field potential
becomes temperature-dependent when quantum corrections are taken into
account \cite{Linde}. Generalization of Coleman results to the case of
nonzero temperature is based on the remarkable fact that quantum statistics
at nonzero temperature is formally equivalent to quantum field theory in the
euclidean space, which is periodic in time coordinate with the period $%
T^{-1} $. As a result, most likely fluctuations appear to be not $O(4)$
symmetric spherical bubbles but $O(3)$ symmetric (with respect to spatial
coordinates) cylindric configurations with certain critical size slightly
different from $O(4)$ symmetric case \cite{Affl,Linde}.

For bubble created with a size smaller than the critical one, it could seem
that the gain in volume energy cannot compensate for the loss in surface
energy and such the bubbles would have to quickly shrink to nothing.
However, detailed analysis discovered that even in this case effects of
nonlinearity lead to nontrivial dynamics. The evolution of subcritical
bubbles - unstable spherically symmetric solutions of nonlinear Klein-Gordon
equation - was, firstly, studied numerically by Bogolubsky and Makhankov 
\cite{Bogolubsky}. Using a quasiplanar initial configuration for the
bubbles, they found that for a certain range of initial radii, the bubble,
after radiating most of its initial energy, settled into long-lived (as
compared with characteristic time-scale) stage and only then disappeared by
quick radiating their remaining energy. Those configurations called
''pulsons'' were later rediscovered and revised by Gleiser who found that
their most characteristic feature is not pulsating mechanism for radiating
the initial energy, but the rapid oscillations of the amplitude of a scalar
field during long-lived pseudo-stable regime when almost no energy was
radiated away and radial pulsations were rather small \cite{Gleiser}. It was
shown \cite{Gleiser,Copeland} that those configurations called ''oscillons''
exist for symmetric and asymmetric double-well potentials, are stable
against small radial perturbations, and have lifetimes ''far exceeding naive
expectations'' \cite{Copeland}.

Although it is well known, that three-dimensional nontrivial configurations
of a scalar field are unstable, they can be relevant for systems with short
dynamical time-scales. Detailed study of unstable but long-lived
configurations can clarify dynamics of nonlinearities in field theories and
their role in a wide class of phenomena ranging from nonlinear optics to
phase transitions both in the Universe and in the laboratory \cite{Copeland}.

For the bubbles formed with the radii large enough (overcritical bubbles) it
is classically energetically favorable to grow. The newly formed bubble of
true vacuum is separated from the surrounding false vacuum region by the
wall at rest. Immediately after nucleation, the wall starts to accelerate
outwards absorbing energy stored in false vacuum region and converting
difference of false and true vacuum energy density into kinetic energy of
the wall . That way a bubble spreads off converting false vacuum into the
true one. This process continues up to the collision with a spherical wall
of another bubble.

In the first-order phase transitions at the end of inflation the collision
of bubbles is considered as the leading mechanism of reheating by converting
the wall energy into radiation. However situation with two bubbles appears
much more complicated. Even nucleation of two bubbles is not yet studied in
the literature in general \cite{Watkins}. Only in the case when bubbles are
widely separated at the time of nucleation and thus can be treated as
noninteracting (at the stage of nucleation) the generalization of a single
bubble solution is straightforward.

Two bubble collisions were studied in detail by Hawking, Moss, and Stewart 
\cite{HMS} and then by Watkins and Widrow \cite{Watkins}, in elegant
approach using symmetry of the problem in zero temperature case. For zero
temperature bubbles produced by quantum tunneling, initial state is $O(4)$
symmetric, as well as euclidean equation of motion, in natural assumption
that a scalar field $\phi$ is invariant under 4-dimensional Euclidean
rotations. In analytical continuation to Minkowski space this becomes $%
O(3,1) $ symmetry. For two bubbles, the line joining their centers is the
preferred axis and solution to the euclidean equation of motion is found in
the class $O(3)$ ($O(2,1)$ as continued to the Minkowski space) solutions 
\cite{HMS,Watkins}, and field configuration arising in collision belongs to
the class of $O(3)$ symmetric solutions.

The aim of present paper is investigation of two-bubble collision in the
case of finite temperature. We are interested not in reheating by two bubble
collisions \cite{Guth} but in evolution of two bubble configuration during
and after collision. The similar subject was discussed several years ago 
\cite{Tkachev}. As was noted by Hawking, Moss and Stewart and confirmed by
Watkins and Widrow, collision of two domain walls does not lead to immediate
convertion of the wall energy into a burst of radiation. Two walls reflect
off one another and move apart creating a new region of false vacuum between
them \cite{Watkins}. Our aim was to investigate an evolution of this new
false vacuum region to look if it can form a separated object. We connected
with such a possibility the hope of formation of metastable relics of the
first order phase transitions such as primordial black holes or
selfgravitating particlelike structures with de Sitter-like cores \cite
{Dymn0}.

To our surprise, a false vacuum configuration evolved into a compact
quasilump filled with an oscillating scalar field. The fundamental
difference of this object from an oscillon is that it arises dynamically as
the result of bubble collisions (which increases probability of its
production) and that it is made up from an oscillating scalar field at the
background of true vacuum. We call it quasilump, since it does not satisfy
all requirements for lumps as defined by S. Coleman: "Non-singular,
non-dissipative solutions of finite energy, lumps of energy holding
themselves together by their own self-interactions" \cite{lump}. In our
numerical simulations we observe non-singular configurations of
self-interacting scalar field with asymmetric potential, perfectly
localized, but we cannot say that they are non-dissipative, although they
are rather long-lived as compared with the characterstic scale for the first
order phase transitions.

Our paper is organized as follows. In Sect.2 we present basic equations and
initial configuration. In Sect.3 we give qualitative analysis of the process
of collision to find an optimal range of parameters for which concentration
of false vacuum energy (which is later transformed into the energy of
oscillations) is maximal. In Sect 4 we present the results of numerical
simulations. Sect 5 contains summary and discussion.

\section{Basic equations}

To study mechanism of formation and evolution of false vacuum regions, we
shall consider most favorable regime for their appearance which corresponds
to high nucleation rate $NH^4\gg 1$ where $N$ is the nucleation rate per
unit four-volume and $H$ is the Hubble parameter\cite{HMS}. We also neglect
gravity effects on the process of bubble formation and growth which means
that we consider bubbles with the initial size much less than cosmological
horizon, $R(0)H\ll 1$ \cite{Coleman,Watkins}.

We consider real scalar field $\phi$ with the Lagrangian

\begin{equation}  \label{Lagr}
{\cal L}=\frac 12\partial _\mu \phi \partial ^\mu \phi -V(\phi).
\end{equation}
This is the effective Lagrangian for a large number of more complex models
of Universe involving the first order phase transitions (see \cite{Linde}
for more details). In the "thin wall" approximation, $\epsilon $$/\lambda
<<1,$ some analytical results are known \cite{Coleman}, and we will work in
the frame of this approximation.

To compare our results with the results obtained by Hawking, Moss and
Stewart, and by Watkins and Widrow, we choose asymmetric double-well
potential of the same form

\begin{equation}  \label{Pot}
\begin{array}{c}
V(\phi )=\frac 18\lambda (\phi ^2-\phi _0^2)^2+\epsilon \phi _0^3(\phi +\phi
_0).
\end{array}
\end{equation}

At $T=0$ the parameters $\lambda,\phi_0$ and $\epsilon$ are specified by the
particle model. At nonzero temperature they are influenced by temperature
corrections. In the case of high nucleation rate, a first order phase
transition is a quick process and we can consider parameters as $\lambda
\simeq $$\lambda (T_{c}), $ $\phi _0\simeq \phi _0(T_{c}),$ $\epsilon \simeq
\epsilon $$(T_{c})$, i.e.being constant during the phase transition at the
temperature $T=T_{c}$.

The potential (\ref{Pot}) has two minima at different values of field $\phi $
. False vacuum (metastable) state is characterized by the field $\phi =\phi
_0(1-\epsilon /\lambda-3/2(\epsilon/\lambda)^2)$, whereas the global minimum
of the potential $V(\phi )$ represents the true vacuum state $\phi
=-\phi_0(1-\epsilon /\lambda +3/2(\epsilon/\lambda)^2)$.

In our analysis we assume that both mechanisms of the false vacuum decay
could take place - tunneling, that is creation of O(4) symmetrical bubbles,
and formation of O(3) symmetrical bubbles due to temperature fluctuation.
Evidently, if the temperature is small enough, the tunneling mechanism of
the false vacuum decay dominate. On the contrary, at large temperatures the
decay is realized by the nucleation and growth of the O(3) symmetrical
bubbles.

Consider conditions of dominance of the false vacuum decay due to
temperature effects. The temperature decay probability was found in \cite
{Affl}, \cite{Linde}:

\begin{equation}  \label{Ptemp}
P_{Temp}\propto e^{-S_3/T},
\end{equation}
where $T$ is the temperature of a phase transition and $S_3$ is
three-dimensional action for O(3) symmetrical bubble. The probability of the
vacuum decay due to tunneling, is given by

\begin{equation}  \label{Ptun}
P_{tun}\propto e^{-S_4}
\end{equation}
where $S_4$ is the action for O(4) symmetrical bubble. The temperature decay
dominates if $S_3/T<S_4.$

The straightforward calculations of the actions $S_3$ and $S_4$ give for our
potential the condition for the dominance of the temperature decay (the term
proportional to $(\epsilon /\lambda )^2$ was omitted):

\begin{equation}  \label{one}
T>\frac{32}{27\pi }\frac \epsilon \lambda
\end{equation}
in the units $m_\varphi =1.$

Consider the equation of motion of the scalar field in spherical coordinates

\begin{equation}  \label{geneq}
\frac{\partial ^2\phi }{\partial t^2}-\frac{\partial ^2\phi }{\partial r^2}%
-\frac 2r\frac{\partial \phi }{\partial r}=-V^{\prime }(\phi ),
\end{equation}

Neglecting terms of order of O($(\epsilon /\lambda )^2$ ), we obtain the
well known one-dimensional equation

\begin{equation}  \label{kink}
\frac{d^2\phi }{dt^2}-\frac{d^2\phi }{dr^2}=-V^{\prime }(\phi )\mid
_{\epsilon =0}.
\end{equation}
The properties of this equation have been extensively discussed in the
literature since 1975 \cite{Goldstone}. The fundamental time independent
solution is defined by 
\[
r=\int_{0}^{\phi}{\frac{d\phi}{\sqrt{2V(\phi)}}}. 
\]
It can be easily checked by straight substitution, that for the theory
defined by potential (\ref{Pot}), the solution is represented in the form

\begin{equation}  \label{solution}
\phi =\phi _0\{th[\frac{\gamma m}2(r-R(t))]-\epsilon /\lambda \},R(t)=vt+R_0,
\end{equation}
where $\gamma =1/\sqrt{1-v^2},$ $v<1$, $m=\sqrt{\lambda}\phi _0$ and $%
R_0=2\lambda /(3\epsilon m)$ is critical radius of the nucleated bubble.

The initial field configuration can be defined at the moment of the bubble
formation $t=0$ with velocity $v=0$. But it takes too much computer time and
memory to obtain the results of the collision, because the kinetic energy of
the walls of the colliding bubbles should be large enough to produce false
vacuum bag (FVB) and hence the initial distance between the centers of
colliding bubbles should be large comparing with critical radius $R_0$ as
well. So, we have to use the initial configuration with already moving walls.

The one-bubble solution (\ref{solution}) is the approximate solution to
exact Eq.(\ref{geneq}). It also satisfies the correct boundary conditions at
infinity up to the terms of order $(\epsilon /\lambda )^2$ and hence can be
chosen as new initial condition at definite moment $t$ or at definite radius
of the expanding bubble $R=R(t)$. The only thing that remains to do is to
connect the radius $R$ and the velocity $v.$

To find the velocity $v$ in the one-bubble solution (\ref{solution}) at an
arbitrary moment $t$ or at definite bubble radius $R(t)$ we note that the
energy

\[
E=\int \biggl(\frac 12(\frac{d\phi }{dt})^2 +\frac 12(\nabla \phi )^2+V(\phi
)\biggr)d^3x 
\]
is conserved if the field $\phi $ is governed by Eq.(\ref{geneq}). The
substitution of the field $\phi $ in form (\ref{solution}) leads after
simple calculations to the expression 
\begin{equation}  \label{HeqConst}
E\simeq \frac{8\pi }3\frac 1\lambda R\left( t\right) ^2[\gamma -R\left(
t\right) \epsilon /\lambda ]=Const
\end{equation}

The $Const$ can be determined at $t=0,$ because we know the values of the
parameters at this moment: $\gamma (t=0)=1$ and $R(t=0)=2\lambda/3\epsilon$ 
\cite{Linde}. Substituting it into expression (\ref{HeqConst}), we find the
connection between the bubble radius $R=R(t)$ and $\gamma -$factor (or,
equivalently, the velocity $v$):

\begin{equation}  \label{gamR}
\gamma =R\frac \epsilon \lambda +\frac 4{27}\frac{\lambda ^2}{\epsilon ^2R^2}
\end{equation}

Thus, the initial conditions for one bubble of radius $R$ is represented by
formula (\ref{solution}) with $\gamma -$factor (\ref{gamR}).

\section{Bubble collisions - Qualitative analysis}

Let us introduce the dimensionless variables $\psi =\phi /\phi _0$, $\lambda
^{1/2}\phi _0t\rightarrow t$ and $\lambda ^{1/2}\phi _0{\bf r}\rightarrow 
{\bf r}$. The classical equation of motion for the scalar field of
Lagrangian (\ref{Lagr}) has the form

\begin{equation}  \label{equ}
\partial _t^2\psi -\nabla ^2\psi =-\frac 12\psi (\psi ^2-1)-\epsilon /\lambda
\end{equation}
The suitable initial two-bubble configuration has in our dimensionless
variables the form

\begin{equation}  \label{init3}
\begin{array}{c}
\psi =\psi _0\{th[\frac \gamma 2(r_{+}-R)]-\epsilon /\lambda \},z<0, \\ 
\psi =\psi _0\{th[\frac \gamma 2(r_{-}-R)]-\epsilon /\lambda \},z>0, \\ 
r_{\pm }=\sqrt{x^2+y^2+\left( z\pm b\right) ^2}, b>R.
\end{array}
\end{equation}

To start numerical simulations of bubble collisions, we need a set of simple
criteria indicating a proper range of parameters favorable to formation of
separated false vacuum regions.

Let us first find the condition at which the region of a false vacuum can be
formed as a result of a collision of two relativistic bubbles.

A field configuration in a wall is just transition from a true vacuum inside
to a false vacuum outside. While propagating through a false vacuum before
collision, bubble absorbs the energy of a surrounding false vacuum and
transforms it into a kinetic energy of the wall. The kinetic energy is
characterized by the Lorentz factor $\gamma=1/\sqrt{1-v^2}$. To get a region
of a false vacuum between bubbles as a result of a collision, energy
absorbed by walls from a false vacuum to the moment of a collision, must be
sufficient to form a false vacuum state at least at the scale of the wall
width. Let us estimate the lower limit for $\gamma$ at which such minimal
region can be formed.

Consider collision of two sphericall $O(3)$ bubble walls described by the
solution (\ref{solution}) with the parameters $R$ and $\gamma_{in}$ to the
moment of a collision. The leading term in the energy density of a wall, as
calculated for the quasiplanar solution (\ref{solution}), is 
\[
\rho_w\simeq{\gamma}^2/4\cosh^4{(\gamma(R-vt)/2)} 
\]
. Before a collision, in the solid angle 
\[
\Delta\Omega=\frac{\pi r^2}{R^2}\ll 1 
\]
, each wall has the energy 
\[
E_{in}=\frac{2}{3}\Delta\Omega R^2 \gamma_{in}. 
\]
After a collision, the walls reflect with a final kinetic energy $E_{fin}$.
If we want bubble wall collision to form, between reflecting walls, a false
vacuum region of a radius $r$ and width $h$ within the solid angle $%
\Delta\Omega$, we must have 
\[
E_{in}=E_{fin}=\frac{2}{3}\Delta\Omega R^2\gamma_{fin} +2\frac{\epsilon}{%
\lambda} V_{fvr}, 
\]
where $\rho_{vac}=2\epsilon/\lambda$ is false vacuum density for the case of
the potential (\ref{Pot}), and $V_{fvr}$ is the volume of a false vacuum
region within a cone with a solid angle $\Delta\Omega$ which is given by 
\[
V_{fvr}=\frac{\pi}{6}h(3r^2+h^2)\approx{\frac{1}{2} \Delta\Omega R^2 h} 
\]
The width of a false vacuum region is of order of a width of a wall to the
moment of reflection which is $2/\gamma_{fin}$. For $\gamma_{fin}=1$, the
width is $h=2$. It gives us the constraint for $\gamma_{in}$ with which a
wall came to the first collision in the form 
\begin{equation}  \label{gamin}
\gamma_{in}\geq 1 + \frac{3}{2}\frac{\epsilon}{\lambda} h \geq 1+3\frac{%
\epsilon}{\lambda}.
\end{equation}
Now let us specify the line joining centers of bubbles as $z$ axis. Let us
show that the energy conservation puts constraint on the propagation of a
false vacuum region in $z$ direction.

Consider a slice of a false vacuum region originated from the collision in
the element $\Delta\Omega$ of spherical bubble walls which has radius $R$ in
the moment of collision. The acceleration of the considered element of the
wall came from transformation of the energy of surrounding false vacuum into
kinetic energy of the wall on the way to its first collision, when the true
vacuum bubbles grow from the initial radius $R(0)$ to the radius $R\gg R(0)$
in the moment of collision. So, the kinetic energy absorbed by the wall from
a false vacuum to the moment of collision, is 
\[
E_{kin}=\frac{2\epsilon}{\lambda}\Delta\Omega\frac{1}{3}(R^3-R(0)^3) \simeq{%
\frac{2\epsilon}{\lambda}\Delta\Omega\frac{1}{3}R^3}. 
\]
The walls reflect each other in the moment of the first collision and move
outwards, creating a false vacuum region between them. Each wall stops when
all its kinetic energy has been transformed into the energy of a false
vacuum region formed between the walls. In this moment the walls radius is $%
R_{max}$ and a false vacuum, created by each wall, fills a region between
the spherical shells $R_{max}$ and $R$. The energy balance gives 
\[
\frac{2\epsilon}{\lambda}\Delta\Omega\frac{1}{3}R^3\simeq{\ \frac{2\epsilon}{%
\lambda}\Delta\Omega\frac{1}{3}(R_{max}^3-R^3)}, 
\]
so that 
\begin{equation}  \label{rmax}
R_{max}\simeq{2^{1/3}}R.
\end{equation}
Since we consider overcritical bubbles, the wall surface energy is neglected
in this treatment, provided that $\epsilon R_{max}/\lambda\gg 1$ The same
result has been obtained for the case of $O(4)$ symmetric bubbles by HMS 
\cite{HMS}.

One finds from the equation (\ref{rmax}) that after the collision a false
vacuum is formed and occupates a region between the outgoing walls, with a
maximal size given by a distance between the planes $z=\pm(2^{1/3}-1)b$,
where $2b$ is the initial separation of the centers of true vacuum bubbles.

After the walls stop their outward movement at $R_{max}=\pm (2^{1/3}-1)R$ in
the region of walls intersection, the parts of walls in this region reflect
off one another and next time they collide at $\Delta t\sim{2(2^{1/3}-1)R}$
after the first collision. The shortest interval between two subsequent
collisions is at $r=0$ when $R=b$. Using the condition (\ref{gamin}), we
find from the Equation (\ref{gamR}) the minimal $\gamma$ at which a false
vacuum region is formed between the walls after the second collision. Before
the second collision at $r=0$, the value of $\gamma$ in the walls is given
by 
\begin{equation}  \label{gam2}
\gamma_2=R\frac{\epsilon}{\lambda} \approx{(2^{1/3}-1)b\frac{\epsilon}{%
\lambda}}.
\end{equation}
Remember that before the first collision this factor for the walls at $r=0$
is given by the Eq.(\ref{gamR}) as 
\begin{equation}  \label{gam1}
\gamma_1=b\frac{\epsilon}{\lambda}.
\end{equation}
If we want the false vacuum region be maintained after the second collision
of the reflected parts of walls, we must satisfy $\gamma_2>\gamma_{in}$.
Then it follows from the Equations (\ref{gamin},\ref{gam2},\ref{gam1}), that 
$\gamma$ before the first collision must be 
\begin{equation}  \label{gammin}
\gamma_1>\gamma_{min}=\frac{1+3\epsilon/\lambda}{(2^{1/3}-1)}.
\end{equation}
It indicates the favorable range of the $\gamma$ parameter before the first
collision needed for numerical simulation, and also, by (\ref{gamR},\ref
{gam1}), the favorable range for the parameters $R$ and $b$.

In the case $\gamma_1\gg \gamma_{min}$, a false vacuum region undergoes the
succession of oscillations - expansions and contractions - along the $z$
axis in the region confined by 
\begin{equation}  \label{planes}
-b(2^{1/3}+1)<z<b(2^{1/3}-1).
\end{equation}

Repeating the above reasoning for the subsequent collisions we find easily
that in the limit of large $\gamma$ the period of the $n-$th oscillation
decreases as $(\sqrt{2n})^{-1}$. This agrees with the HMS result \cite{HMS}
for the $O(4)$ symmetry case. The reason for such a coincidence can be
easily understood. The main difference between the $O(3)$ and $O(4)$ cases
is in the form of the initial wall configurations, taken in our case as a
quasiplanar $O(3)$ solution. However, in all the above reasoning the
internal structure of the walls was not involved which just resulted in
similar estimation for the decreasing of period of oscillations.

For large $\gamma$ we can treat the oscillations of a false vacuum region
along the axis $z$ as the continuous propagation of a spherical wave moving
with speed of light (in our units $c=1$). In the frame with the origin in,
say, $z=-b$, the element $\Delta\Omega$ of the wall with the angle $\alpha$
with respect to the $z$ axis, follows the trajectory $r=z \tan{\alpha}$.
Assume that to the moment of reflection considered element of the wall has
the coordinate $z=ct$. At the same moment its radial coordinate is $r=z\tan{%
\alpha}$. The region of causal contact along the axis $r$ satisfies the
condition $r\leq ct$. It follows then that only for the angles $\alpha<\pi/4$%
, the region of intersection of walls is in causal contact. Therefore the
boundary of the region of causal contact within a false vacuum region is the
cone 
\begin{equation}  \label{45}
\alpha=\pi/4.
\end{equation}
It means that further evolution of a false vacuum region confined within the
boundary (\ref{45}) does occur independently on dynamics of field outside
this boundary. So, considered region of a false vacuum is separated in its
further causal and hence dynamical evolution.

Now we can easily estimate the total energy of a separated false vacuum
region. The energy density of a false vacuum is given by $%
\rho_{vac}=2\epsilon/\lambda$. The volume of a FVB is the volume of two
cones whose height is equal $b$ and base area $\pi b^2$. So, the mass
confined within this region is 
\begin{equation}  \label{massFVB}
M=\frac{4\epsilon}{\lambda}\frac{\pi b^3}{3}.
\end{equation}
It is evident that separation occurs at the time of order of $t_{sep}=(\sqrt{%
2}-1)b$.

\section{Numerical Results}

In the cylindric coordinates the equation of motion for a scalar field (\ref
{equ}) equation takes the form 
\[
\partial_t^2\psi - \partial_r^2\psi - \partial_z^2\psi =-\frac 12\psi (\psi
^2-1)-\epsilon /\lambda 
\]
The solution to this equation has the axial symmetry and reflection symmetry
with respect to $z=0$ plane. The initial configuration described by
solutions 
\begin{equation}  \label{psi}
\psi=th[\gamma/2(r_{+}-R-vt)]+th[\gamma/2(r_{-}-R-vt)]-1-\epsilon/\lambda
\end{equation}
is shown in Fig.1. The walls already have kinetic energy that is indicated
by $\gamma$ factor equals $5$.


Time evolution of a scalar field in the center of the region of collision $%
\psi (t,r=z=0)$ shown in Fig.2, was calculated for the parameters $\Gamma
=5;b=52;R=50$. The qualitative behavior of the field with time has been
discussed in previous section. From the beginning the field changes in the
manner discussed in \cite{HMS}, \cite{Watkins}, then it oscillates around
true vacuum for a long time and finally, large secondary fluctuations appear
again. \vskip 0.1in


As we shall see below, the energy of oscillations is perfectly localized.
This behaviour does not change with changing the step in calculations. Field
configurations in different moments of time are shown in Figs.3,4.



The energy density profile calculated from scalar field potential, is shown
in the next series of figures. They demonstrate concentration of field
energy in the center of region of collision. In the fig.5 one can see time
dependence of energy density in the center of the region of collision. Large
secondary peak is created due to the coherent field oscillation which are
coming from outside. 

The density profile at this time is represented in the Fig.6. This localized
configuration oscillates for some time and finally is converted into
outgoing radiation. Only gravity could prevent this process. Till now we did
not consider gravitational effects, but they will be estimated below. 
Consider the evolution of energy contained in the sphere of certain radius
as shown in Fig.7-8.



This pictures show two peaks of energy - the first is due to energy in the
moment of collision, the second is the energy of the quasilump which is
formed as a result of the collision. It becomes evident by comparing these
figures that the energy is strongly localized. Indeed, the energy contained
in internal sphere of radius 5 is only in 1.5 times smaller than that
contained in external sphere, while the ratio of their volumes is 8.
Gravitational radius of the quasilump in our units is

\begin{equation}
r_g=\left( {\frac{m}{m_{pl}}}\right) ^2E/\lambda  \label{rg}
\end{equation}

If the first order phase transition happens at the end of inflation, the
mass $m$ of inflaton field is rather large and ${m_\varphi /m}_{pl}\sim
10^{-5}$. Substituting this value into (\ref{rg}) and the value of energy $%
E\approx 1000$ obtained from fig.7 one can easily find the condition when
gravitational radius is comparable with the size of the quasilump, which can
be taken as $r_0 \approx 5$in our dimensionless units. This condition is
satisfied if coupling constant $\lambda \sim 10^{-8}$. For $\lambda <10^{-8}$
gravitational forces become essential and the probability of black hole
formation grows up to unity when $\lambda $ tends to $10^{-8}$.

\section{Summary}

In this paper we give qualitative arguments supported by numerical
simulation for the existence of long-lived fluctuation that arise as a
result of a collision of two expanded bubbles.

The two-bubble collision leads, first, to the formation of short-living
false vacuum region in the center of collision. Numerical results indicate
separation of a false vacuum region at the time $t\sim b$. Then it evolves
into rather compact object - quasilump made up of a scalar field oscillating
around its true minimum, with lifetime enough to be captured by its
gravitational field. At small coupling constants black hole can be produced.

Till now the similar object discussed in literature was oscillon \cite
{gleiser1}. The main difference between these two objects is as follows. i)
Oscillon represent a subcritical bubble of true vacuum inside a false
vacuum, that arise due to temperature fluctuations. Our object is the
fluctuation of scalar field in the true vacuum background that arise as a
result of dynamical process. ii) To be long-lived, oscillon should have
rather large initial radius, though less than critical one, and rather flat
initial distribution of scalar field. The evolution of the oscillon consists
of oscillation of field value with almost constant radius of the field
configuration. Our lump is much more compact object with the amplitude value
of scalar field being much larger than that of the field in its potential
minimum. For $\lambda \leq 10^{-8}$ gravitational forces are essential and
the probability of PBH creation is of order unity. iii) Oscillon, being
produced in spite of small probability \cite{riotto}, is extremely long
lived object with lifetime 103 -104 1/m, m being the mass of scalar field.
The life-time of our lump is of the same order of magnitude but lumps can be
produced with much heigher probability, because they result from collisions
of overcritical bubbles whose rate of nucleation is much bigger than for
subcritical bubbles.

\section{Acknowledgement}

This work was supported by the Polish Committee for Scientific Research
through the Grant 2P03D.017.11. The work of MyuK and SGR was partially
performed in the framework of Section "Cosmoparticle physics" of Russian
State Scientific Technological Program "Astronomy. Fundamental Space
Research", 
with teh support of Cosmion-ETHZ and Epcos-AMS collaborations.

\end{document}